\documentclass[twocolumn,showpacs]{revtex4}

\usepackage{bm}

\usepackage{graphicx}
\usepackage{subfigure}

\begin{document}

{\large\bf Entanglement and teleportation of macroscopic continuous
variables by superconducting devices}

\bigskip{\normalsize\bf
L.~Fisch$^1$, I.E.~Mazets$^1$\,$^,$\,$^2$, and G.~Kurizki$^1$ }

\medskip\noindent{\small\it $^1$\,Department of Chemical Physics, Weizmann
Institute of Science, Rehovot 76100, Israel, \\
$^2$\,Ioffe Physico-Technical Institute, St.Petersburg 194021,
Russia \\}

\dotfill

\noindent {\bf A current-biased low-temperature superconducting
Josephson junction (JJ) is dynamically describable by the quantized
motion of a fictitious particle in a ``washboard'' potential$^1$.
The long coherence time of tightly-bound states in the washboard
potential of a JJ has prompted the effort$^{2-6}$ to couple JJs and
operate them as entangled qubits, capable of forming building blocks
of a scalable quantum computer.  Here we consider a hitherto
unexplored quantum aspect of coupled JJs: the ability to produce
Einstein-Podolsky-Rosen (EPR) entanglement of their continuous
variables$^7$, namely, their magnetic fluxes and induced charges.
Such entanglement, apart from its conceptual novelty, is the
prerequisite for a far-reaching goal: teleportation$^{8-11}$ of the
flux and charge variables between JJs, implementing the transfer of
an unknown quantum state along a
network of such devices. \\[1eM]}

For our analysis, we shall adopt the following Hamiltonian$^5$ of
two JJs coupled by capacitance $C_c$, written in terms of the
quantized variables $\theta _{1,2}=\Phi _{1,2}/\Phi _0$, i.e., their
magnetic fluxes normalized to the flux quantum $\Phi _0= 2\pi \hbar
/(2e)$, and their canonically-conjugate variables $p_{1,2}=
-i\partial /(\partial \theta _{1,2})$, related to induced charges:
\begin{eqnarray}
H&=&4E_C(1+\zeta )^{-1} (p_1^2+p_2^2+2\zeta p_1p_2) \nonumber \\ &&
-E_J(\cos \theta _1+\cos \theta _2+J_1\theta _1+J_2\theta _2).
\label{eq:1}
\end{eqnarray}
Here $E_C=e^2/C_J$, $C_J$ being the single-junction capacitance;
$E_J=\hbar I_c/(2e)$, $I_c$ being the critical current; $J_{1,2}=
I_{1,2}/I_c$ are the normalized bias currents, and $\zeta
=C_c/(C_c+C_J)$ is the coupling parameter.

In order to study EPR correlations, we adopt the basis of collective
variables
\begin{equation}
p_\pm =(p_1\pm p_2)/\sqrt{2}, \quad \theta _\pm =(\theta _1\pm
\theta _2)/\sqrt{2}, \label{eq:2}
\end{equation}
thereby rewriting Eq. (\ref{eq:1}) as
\begin{eqnarray}
H&=&\frac {p_+^2}{2m_+}+\frac {p_-^2}{2m_-} -\frac
{E_J}{\sqrt{2}}(J_1+J_2)\theta _+ \nonumber \\ && -\frac
{E_J}{\sqrt{2}}(J_1-J_2)\theta _- -2E_J\cos \frac {\theta
_+}{\sqrt{2}}\cos \frac {\theta _-}{\sqrt{2}}. \label{eq:3}
\end{eqnarray}
Here the + and -- collective modes are characterized by different
``masses'' $m_\pm =(1+\zeta )/[8E_C(1\pm \zeta )]$. The coupling of
$\theta _+$ and $\theta _-$ via the last term in Eq. (\ref{eq:3})
renders the dynamics two-dimensional (2D) in the + and -- modes.

EPR correlations occur only between two {\em commuting} variables,
e.g., $\theta _+$ and $p_-$ or $\theta _-$ and $p_+$$^{7, 8, 12}$.
Ideal EPR correlations require these variables to be dynamically
decoupled$^{7, 8, 11, 12}$. Such decoupling of the + and -- modes
holds when the Born-Oppenheimer (BO) approximation$^{13}$, which is
widely used in molecular dynamics, is valid. To impose the BO
approximation, we draw a parallel between the -- and + modes of the
coupled JJs, and, respectively, slow (nuclear) and fast (electronic)
degrees of freedom  of a molecule. This parallel holds for a large
difference of the corresponding masses
\begin{equation}
m_+ \ll m_-    \, .               \label{eq:4}
\end{equation}

Condition (\ref{eq:4}), which justifies the BO decoupling of the
fast and slow modes$^{13}$, means that, during the oscillation of
the $\theta _-$-mode, the $\theta _+$-mode always remains on the
same potential curve. Mathematically, the BO approximation implies
that the eigenfunctions of Eq.~(\ref{eq:3}) are factorizable as
$\varphi (\theta _+)\Phi (\theta _-)$, with the fast-mode
eigenfunctions $\{ \varphi _n(\theta _+) \} $ and the corresponding
energy eigenvalues $\{ \varepsilon _n\} $, $n=0,1,2, \, ... \, $,
being parametrically dependent on the slow-mode coordinate $\theta
_-$.

Let us first consider the unbiased case, $J_{1,2}=0$. Then, to
zeroth order in $m_+/m_-$, the bound-state energies $\varepsilon
(\theta _-^{(0)})$ of the fast mode are minimized for $\theta
_-^{(0)}=0$. They are the Mathieu (pendulum) equation eigenvalues,
whose form for $E_J\gg 1/m_+=8E_C$ (the ``phase regime''$^{2-6}$) is
weakly anharmonic:
\begin{eqnarray}
\varepsilon _n(0)&\approx &-2E_J+\hbar \omega _0\left( n+\frac
12\right)
-\frac 1{16m_+}\left( n^2+n+\frac 12\right) ,\nonumber \\
\hbar \omega _0 &=&\sqrt{E_J/m_+}. \label{eq:5bis}
\end{eqnarray}

\begin{figure}[tbh]
\centering \centering
\scalebox{0.4}{\includegraphics{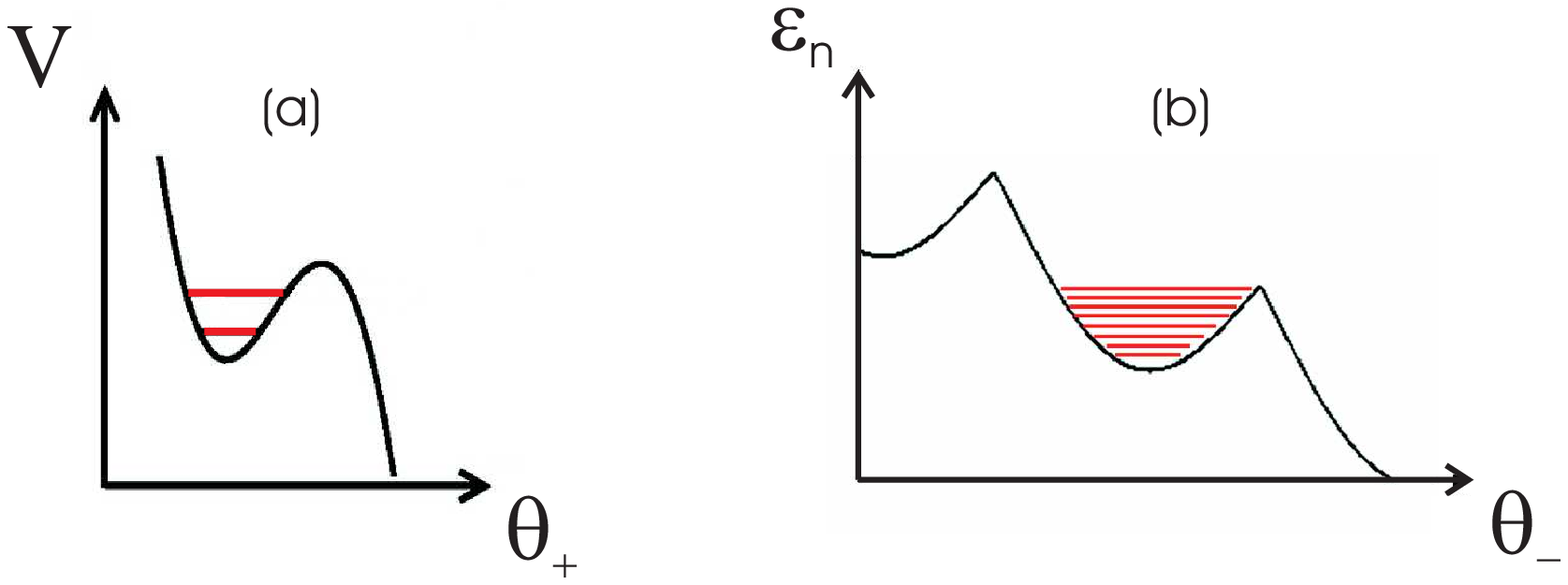}} \centering
\scalebox{0.4}{\includegraphics{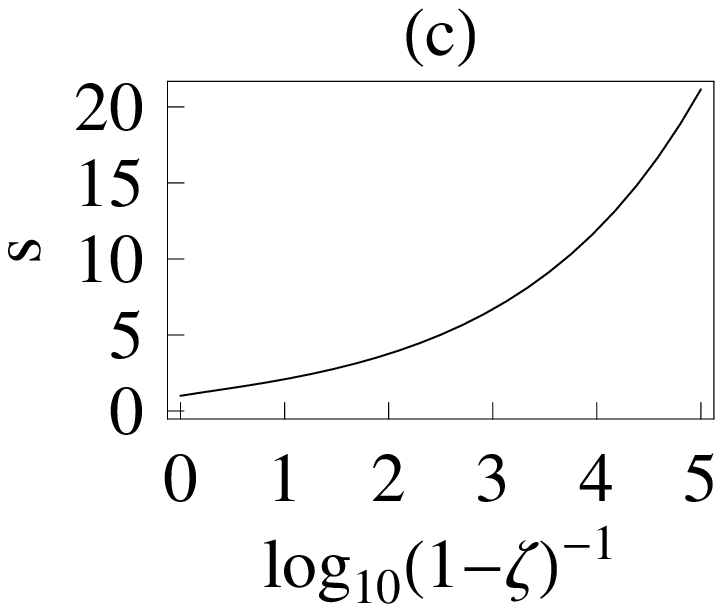}\includegraphics{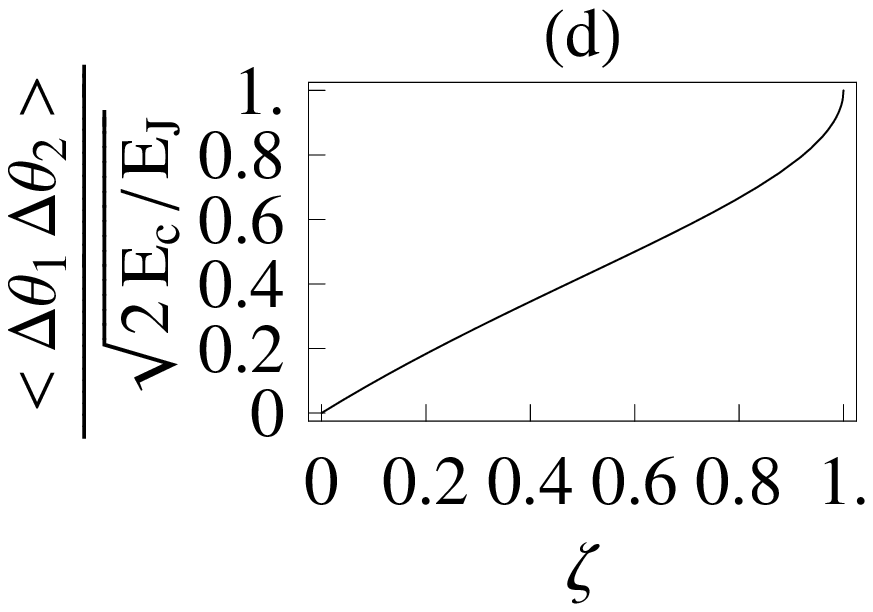}}
\caption{(a) Fast-mode washboard potential with few bound levels,
large tunneling rate for excited states, (b) slow-mode washboard
potential with many energy levels and possibility of
{\emph{quasiclassical}} motion, (c) squeezing $s$ and (d) normalized
cross-correlation $(E_j/(2E_c))^{1/2} \langle \Delta \theta _1
\Delta \theta _2 \rangle$ for the ground state, plotted versus
$\log_{10}(1-\zeta)^{-1}$ (in (c)) or coupling $\zeta $ (in (d)).}
\end{figure}

The bound-state solutions for the slow mode, to first order in
$\sqrt{ m_+/m_- }$, are obtained upon retaining quadratic (harmonic)
terms. The corresponding slow-mode energies, specified by two
quantum numbers ($n$ and $\nu $), are
\begin{eqnarray}
E_{n, \nu }&\approx &
\hbar \Omega _n \left( \nu +1/2 \right) , \label{eq:5PP} \\
\Omega _n&\approx &\omega _0 \left( 1 -\frac {n+1/2}{4\sqrt{E_Jm_+}}
\right) \sqrt{\frac {m_+}{m_-}}. \label{eq:5QQ}
\end{eqnarray}
In the weakly-biased case, $|J_{1,2}|\ll 1$, the ratio of the level
spacings of the slow and fast modes is also of the order of $\sqrt{
m_+ /m_-}$, but the excited levels are somewhat broadened by leakage
to the continuum, [Fig. 1(a,b)].

For the states described by Eqs. (\ref{eq:5bis}\, -- \,
\ref{eq:5QQ}), the signature of EPR correlations (entanglement) is
twofold: (a) The product of {\em commuting-variable} variances of
the collective (+ and --) modes is reduced well below the Heisenberg
uncertainty limit$^{14}$
\begin{equation}
\sqrt {\langle \Delta \theta _\pm ^2\rangle _{n,\nu } \langle \Delta
p _\mp ^2\rangle _{n,\nu }} \equiv \frac 1{2s} \ll \frac 12,
\label{eq:6}
\end{equation}
implying that the variables $\theta _-$ and $p_+$ or their
conjugates become ``squeezed''$^{14}$. Appreciable ``squeezing'',
$s>3$, is achievable for $\zeta > 0.976$. (b) The cross-terms
$\langle \Delta \theta _1 \Delta \theta _2\rangle _{n,\nu }$ and
$\langle \Delta p _1 \Delta p _2\rangle _{n,\nu }$ in $\langle
\Delta \theta _- ^2\rangle _{n,\nu }$ and $\langle \Delta p_+
^2\rangle _{n,\nu }$, expressing inter-junction voltage and current
correlations, are nonvanishing and tend to grow in absolute value
with the coupling parameter $\zeta $ [Fig.~1(c)].

In order to prepare these EPR-correlated states, we may start from
uncoupled JJ1 and JJ2  ($\zeta =0$), both in the ground state, and
quasi-adiabatically increase $\zeta $. The quasiadiabaticity
confines $\theta _+$ and $\theta _-$ to their respective ground
states, but the ground-state uncertainties in Eq.~(\ref{eq:6})
become progressively ``squeezed''. Quasiadiabatic change of the
coupling  capacitance $C_c$ and, hence, of $\zeta $ may be attained
by gradually switching-on a battery of ferroelectric capacitors
$^{15}$ connected in parallel, much slower than $\Omega _0^{-1}$ but
well within the JJ coherence time.

\begin{figure}
\centering \scalebox{0.4}{\includegraphics{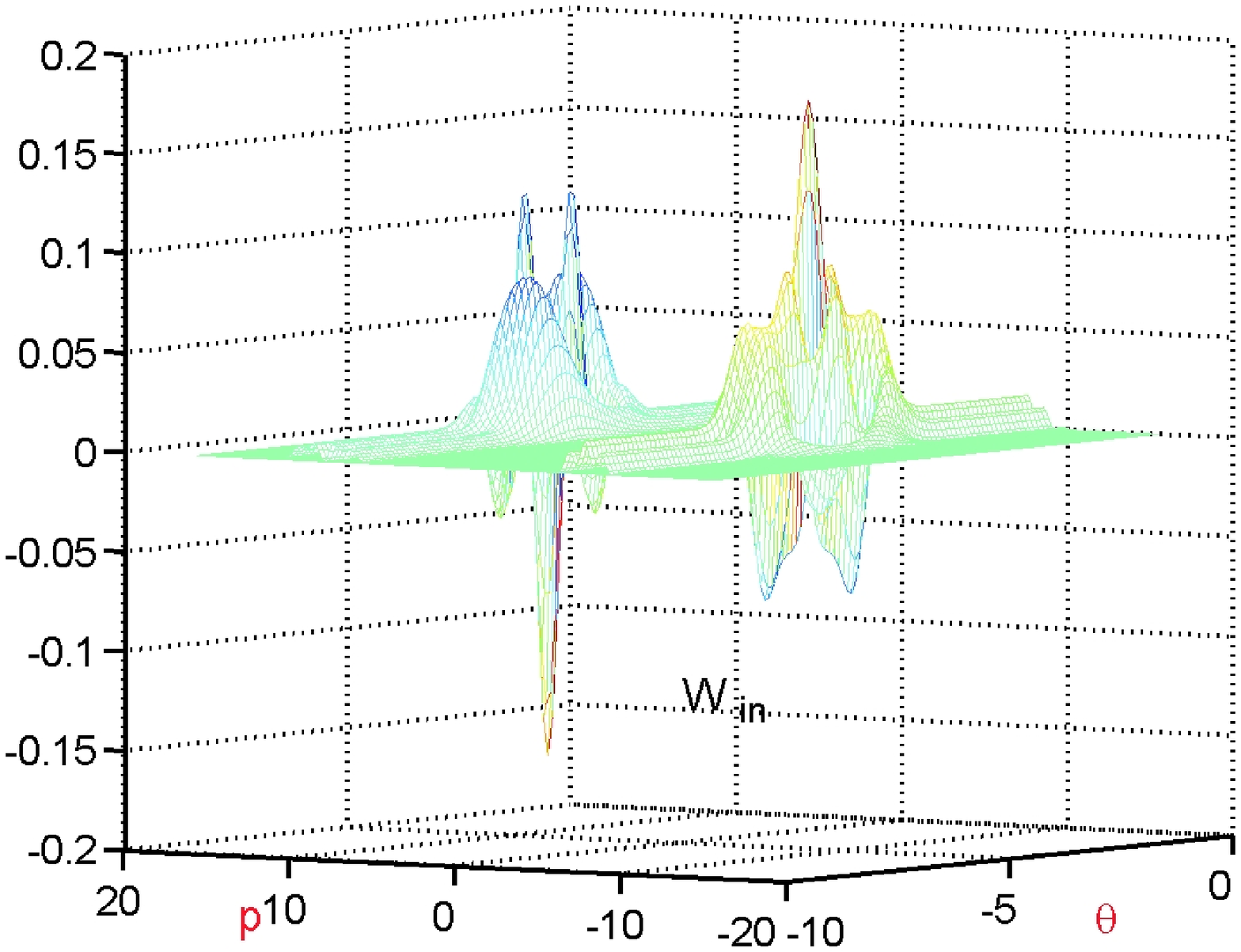}}
\centering \scalebox{0.4}{\includegraphics{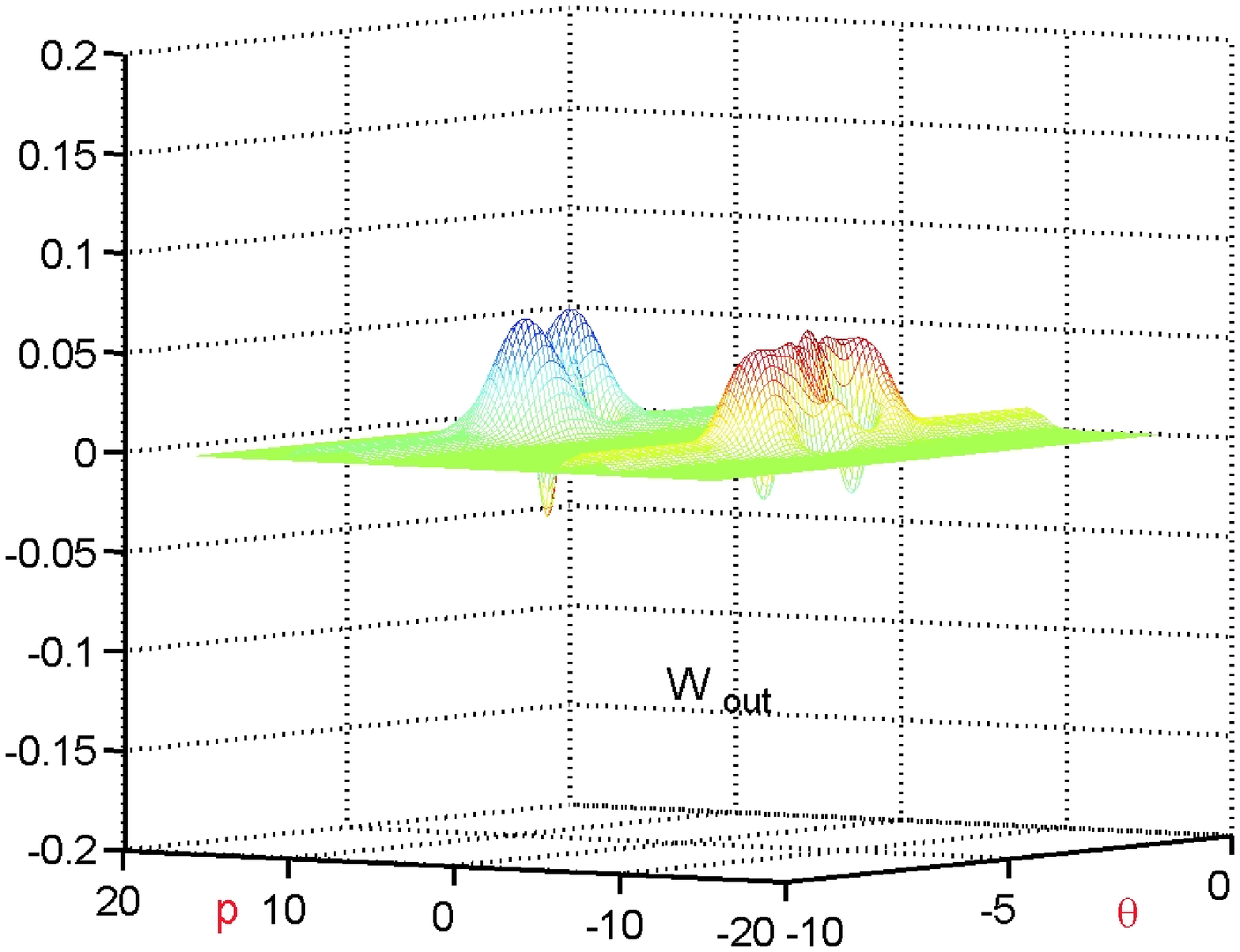}}
\caption{(a) Wigner function for the input states $(|1\rangle -
i|3\rangle)/\sqrt{2}$ (left) and $(|0\rangle - i|2\rangle +
i|4\rangle)/\sqrt{3}$ (right), and (b) corresponding output
(teleported) states, for an effective coupling $\zeta = 0.9995$ (as
per Eq. (11))} \end{figure}

If indeed Eq.~(\ref{eq:6}) is satisfied for a certain interaction
time, as in Fig.~1(c), we may use these EPR correlations for
implementing the continuous-variable teleportation protocol$^{11}$.
To this end, consider JJ3 to be uncoupled from the EPR-correlated
JJs 1 and 2. We wish to teleport the unknown state $|\psi _3\rangle
$ of JJ3 to the distant JJ1. The crucial step of the protocol is the
measurement of the {\em joint variables}
\begin{equation}
\theta _{23}^{(-)}=\theta _2-\theta _3, \quad  p_{23}^{(+)}=p_2+p_3.
\label{eq:7}
\end{equation}
This can be effected by {\em classical probes} that are sensitive to
the sum of fluxes (currents) and the difference of voltages of JJ2
and JJ3.

The next step is the communication of the measurements' results
$\bar{\theta }_{23}^{(-)}$ and $\bar{p}_{23}^{(+)}$ to the
``controller'' of JJ1, who then switches off the coupling $\zeta $
of JJ1 and JJ2 and ``shifts'' the flux and voltage of JJ1:
\begin{equation}
\theta _1\rightarrow \theta _1- \bar{\theta }_{23}^{(-)}, \quad
p_1\rightarrow p_1+\bar{p} _{23}^{(+)}. \label{eq:8}
\end{equation}
Provided that the protocol has been implemented with sufficiently
high fidelity, it transforms the state of JJ1 to replicate $|\psi _3
\rangle $ (see$^{11}$), yielding $\theta _1\rightarrow \theta _3$,
$p_1 \rightarrow p_3$. In Fig.~2 we demonstrate the possibility of
high-fidelity teleportation of a superposition (wavepacket)
consisting of either 3 or 2 ``washboard''-potential eigenfunctions
$\psi _3(\theta) =\sum c_j(0) \varphi _j(\theta ) $ from JJ3 [Fig.
2(a)] to JJ1 [Fig. 2(b)].

The uncertainty (``noise'') that degrades the fidelity of the
teleportation protocol is determined by$^{11, 12}$
\begin{eqnarray}
\langle \Delta \theta _{T}^2 \rangle &=& \langle \Delta \theta
_{12}^{(-)\, 2} \rangle +
\langle \Delta \theta _{23}^{(-)\, 2} \rangle    ,\nonumber \\
\langle \Delta p _{T}^2 \rangle &=& \langle \Delta p _{12}^{(+)\, 2}
\rangle + \langle \Delta p _{23}^{(+)\, 2} \rangle , \label{eq:9}
\end{eqnarray}
where $\langle \Delta \theta _{12}^{(-)\, 2} \rangle $ and $\langle
\Delta p_{12}^{(+)\, 2}\rangle $ are determined by the EPR
correlations discussed above, while $\langle \Delta \theta
_{23}^{(-) \, 2} \rangle $ and $\langle \Delta p _{23}^{(+)\,
2}\rangle $ depend on the accuracy of measuring the joint variables
$\theta _{23}^{(-)}$ and $p_{23}^{(+)}$ and the transfer (shift)
accuracy. High fidelity (defined as the overlap of the input and
output states) requires that $\langle \Delta \theta _{T}^2 \rangle
\langle \Delta p_{T}^2 \rangle \ll 1$. For the conceivable
parameters$^6$ chosen in Fig.~2(b), this ``noise'' entails $\langle
\Delta \theta _T^2 \rangle \simeq 0.015$, $\langle \Delta p_T^2
\rangle \simeq 3.19$, which limits the fidelity of the protocol to
$\lesssim 0.82$.

To conclude, we have demonstrated the feasibility, in principle, of
EPR entanglement and teleportation of macroscopic continuous
variables in superconducting Josephson junctions. Beyond the
innovation of performing quantum operations upon such macroscopic
observables as magnetic flux and induced voltage, their realization
would allow the development of quantum-information transfer networks
comprised of superconducting elements. Such transfer may be used to
interface blocks of quantum-information processors.

{\bf Acknowledgements} We acknowledge the support of the EC (the
QUACS RTN and SCALA NOE), ISF, Russian Leading Scientific Schools
(I.E.M., grant 1115.2003.2), and Universities of Russia (I.E.M.,
grant UR.01.01.287).

\begin{enumerate}

{\small \item Leggett, A.J. {\em et al.} Dynamics of the dissipative
2-state system. {\em Rev. Mod. Phys.} {\bf 59}, 1--85 (1987).

\item Vion, D. {\em et al.} Manipulating the quantum state
of an electric circuit. {\em Science} {\bf 296}, 886--889 (2002).

\item Yu, Y., Han, S.Y., Chu X., Chu S.I., \& Wang Z.
Coherent temporal oscillations of macroscopic quantum states in a
Josephson junction. {\em Science} {\bf 296}, 889--892 (2002).

\item Ioffe, L.B., Geshkenbein, V.B., Feigel'man, M.V.,
Fauchere, A.L., \& Blatter, G. Environmentally decoupled sds-wave
Josephson junctions for quantum computing. {\em Nature} {\bf 398},
679--681 (1999).

\item Berkley, A.J. {\em et al.} Entangled macroscopic
quantum states in two superconducting qubits. {\em Science} {\bf
300}, 1548--1550 (2003).

\item Majer, J.B., Paauw, F.G., ter Haar, A.C.J.,
Harmans, C.J.P.M., \& Mooij, J.E. Spectroscopy on two coupled
superconducting flux qubits. {\em Phys. Rev. Lett.} {\bf 94}, 090501
(2005).

\item Einstein, A., Podolsky, B., \& Rosen, N. Can
quantum-mechanical description of physical reality be considered
complete? {\em Phys. Rev.} {\bf 47}, 777 (1935).

\item Bennett, C.H. {\em et al.} Teleporting an unknown
quantum state via dual classical and Einstein-Podolsky-Rosen
channels. {\em Phys. Rev. Lett.} {\bf 70}, 1895--1899 (1993).

\item  Vaidman, L. Teleportation of quantum states.
{\em Phys. Rev. A} {\bf 49}, 1473--1476 (1994).

\item Bouwmeester, D. {\em et al.} Experimental quantum
teleportation. {\em Nature}  {\bf 390}, 575--579 (1997).

\item Braunstein, S.L. \& Kimble, H.J. Teleportation of
continuous quantum variables. {\em Phys. Rev. Lett.} {\bf 80},
869--872 (1998).

\item Opatrn\'{y}, T. \&  Kurizki, G. Matter-wave entanglement
and teleportation by molecular dissociation and collisions. {\em
Phys. Rev. Lett.} {\bf 86}, 3180--3183 (2001).

\item Davydov, A.S. {\em Quantum Mechanics} (Pergamon, Oxford,
1976), Chapter XV.

\item Scully, M.O. \& Zubairy, M.S. {\em Quantum Optics}
(Cambridge University Press, Cambridge, 1997), Chapter 2.

\item Yoon, Y.K., Kim, D., Allen, M.G., Kenney, J.S., \&
Hunt, A.T. A reduced intermodulation distortion tunable
ferroelectric capacitor --- Architecture and demonstration. {\em
IEEE Trans. Microwave Theor. Techn.} {\bf 51}, 2568--2576 (2003). }

\end{enumerate}


\end{document}